\documentclass[aps,prl,twocolumn,showkeys,groupedaddress,amsmath,amssymb,nofootinbib,showpacs,floatfix]{revtex4-1}
\usepackage{graphicx}  
\usepackage{dcolumn}   
\usepackage{color}
\usepackage{psfrag}
\usepackage{bbm,bm}
\usepackage{epsfig}
\allowdisplaybreaks
\begin{document}
\def\be{\begin{equation}}
\def\ee{\end{equation}}
\def\bea{\begin{eqnarray}}
\def\eea{\end{eqnarray}}
\def\ba{\begin{array}}
\def\ea{\end{array}}
\def\ben{\begin{enumerate}}
\def\een{\end{enumerate}}
\def\nab{\bigtriangledown}
\def\tpi{\tilde\Phi}
\def\nnu{\nonumber}
\newcommand{\eqn}[1]{(\ref{#1})}
\def\bw{\begin{widetext}}
\def\ew{\end{widetext}}
\newcommand{\half}{{\frac{1}{2}}}
\newcommand{\vs}[1]{\vspace{#1 mm}}
\newcommand{\dsl}{\pa \kern-0.5em /} 
\def\a{\alpha}
\def\b{\beta}
\def\g{\gamma}\def\G{\Gamma}
\def\d{\delta}\def\D{\Delta}
\def\ep{\epsilon}
\def\et{\eta}
\def\z{\zeta}
\def\t{\theta}\def\T{\Theta}
\def\l{\lambda}\def\L{\Lambda}
\def\m{\mu}
\def\f{\phi}\def\F{\Phi}
\def\n{\nu}
\def\p{\psi}\def\P{\Psi}
\def\r{\rho}
\def\s{\sigma}\def\S{\Sigma}
\def\ta{\tau}
\def\x{\chi}
\def\o{\omega}\def\O{\Omega}
\def\k{u}
\def\pa {\partial}
\def\ov{\over}
\def\nn{\nonumber\\}
\def\ud{\underline}
\def\ct{\textcolor{red}{\it cite }}



\title{\large{\bf Some aspects of non-perturbative QCD from non-susy D3 brane \\ of Type IIB string theory}}
\author{Kuntal Nayek}
\email{kuntal.nayek@saha.ac.in}
\affiliation{
   Saha Institute of Nuclear Physics,\\
   1/AF Bidhannagar, Kolkata 700064, India\\
   and\\
   Homi Bhabha National Institute,\\
   Training School Complex, Anushakti Nagar, Mumbai 400085, India
   }
\author{Shibaji Roy}
\email{shibaji.roy@saha.ac.in}
\affiliation{
   Saha Institute of Nuclear Physics,\\
   1/AF Bidhannagar, Kolkata 700064, India\\
   and\\
   Homi Bhabha National Institute,\\
   Training School Complex, Anushakti Nagar, Mumbai 400085, India
   }
\date{\today}

\begin{abstract}
It is well-known that the non-supersymmetric D3 brane (a cousin of BPS D3 brane) including its black version 
of type IIB string theory has a decoupling limit, where the decoupled geometry is the gravity dual of a non-supersymmetric, 
non-conformal (finite temperature) quantum field theory having
some properties similar to QCD. Using the ideas of AdS/CFT we study some non-perturbative aspects of this quantum field
theory. Since in this case we have a Yang-Mills theory (no quarks) with running coupling (non-constant dilaton), we compute 
the gluon condensate in this theory as a function of temperature and also compute the beta function. The behavior of the gluon 
condensate is found to 
resemble much like the SU(3) lattice QCD result and the beta function is found to be negative. We further compute both the 
pseudoscalar and the scalar glueball mass spectra in this theory using WKB approximation and 
find that the mass ratios of the first excited state to the ground state of the scalar glueball are quite close to the lattice 
QCD results.         
\end{abstract}

\pacs{11.25.-w, 11.25.Tq, 11.25.Uv}

\maketitle

The non-perturbative QCD is still a very poorly understood regime of QCD as we don't know how to deal with strongly coupled quantum field theories. 
The lattice gauge theory is the only approach to this low energy QCD. However, lattice QCD deals with Euclidean signature and therefore,
is unable to give real time dynamics of the system. Even if we consider only the kinematic properties, there are problems with the lattice
approach, for example, the finite size and spacing of the lattice is a limitation and taking continuum limit is not always easy and computationally
quite challenging. Another approach called the AdS/CFT correspondence \cite{Maldacena:1997re, Aharony:1999ti} has provided a new theoretical tool 
for studying the non-perturbative QCD, 
which is known as holographic QCD or AdS/QCD. Many of the properties of QCD in the low energy has been studied by employing this
approach.   

The AdS/CFT duality, proposed by Maldacena \cite{Maldacena:1997re}, is about the (conjectured) equivalence between string theory in five 
dimensional anti-de Sitter 
(AdS) (times five dimensional sphere) background and the (super)conformal field theory (CFT) in four dimensions. This is a strong-weak 
duality which means that when the field theory is strongly coupled the string theory is weakly coupled, i.e., given by the supergravity and 
{\it vice-versa}. This, therefore, gives a handle on the strongly coupled quantum field theories by studying the dual gravity theories in a 
particular (AdS) background. This procedure has been used to study non-perturbative QCD. However, as the gauge coupling here has to be 
strong enough to get well-defined gravity theory, the asymptotic freedom is absent in this type of quantum field theory \cite{Polchinski:2001tt}. 
But the other properties of low energy QCD like gluon condensate, confinement, chiral symmetry breaking, negativity of beta function etc. 
\cite{Constable:1999ch,Babington:2003vm,Csaki:2006ji} and various properties of QGP 
\cite{Liu:2006he,CasalderreySolana:2011us,Chakraborty:2017wdh} can be understood from gauge/gravity duality. 
The original AdS/CFT \cite{Maldacena:1997re} duality arises from the BPS D$3$ brane solution of type IIB string theory, where the coupling (the dilaton) remains 
fixed and therefore in the dual gauge theory the 't Hooft coupling is large but fixed. Also, because of the presence of conformal symmetry 
this type of gauge theory does not have $\Lambda_{\rm QCD}$-like scale. But the running coupling and 
$\Lambda_{\rm QCD}$ are the two main requirements for QCD. Thus the pure AdS/CFT correspondence (which is also maximally supersymmetric unlike QCD) 
is not the right framework to study QCD.
 
Now to introduce running coupling and $\Lambda_{\rm QCD}$, there are many bottom-up approaches, where people have obtained some 
soft-wall (non-constant dilaton) gravity theories to incorporate features of non-perturbative QCD \cite{Csaki:2006ji,Batell:2008zm}. 
The gravity theories used there are not always obtained as solutions of string theory (in few cases 
\cite{Gubser:1999pk,Kehagias:1999tr,Nojiri:1999uh,Constable:1999ch} they are obtained from string theory). 
They are more like empirical modification of pure AdS background \cite{Brodsky:2010ur,deTeramond:2009xk}.

Here in this Letter we use the gauge/gravity duality on the finite temperature non-susy D3 brane solution 
of type IIB string theory \cite{Zhou:1999nm,Lu:2004ms,Nayek:2015tta,Nayek:2016hsi,Chakraborty:2017wdh}. The 
decoupled geometry, in this case, is non-AdS and non-supersymmetric with a non-trivial dilaton field \cite{Chakraborty:2017wdh}. So the 
expected gauge theory in this duality is non-supersymmetric and non-conformal as in QCD. 
The non-trivial dilaton is a sign of the running coupling, whereas the absence of conformal symmetry indicates the existence of 
$\Lambda_{\rm QCD}$. Here we identify the gauge theory energy scale with the same energy parameter of the gravity theory $u$. Other 
two parameters of gravity background, namely, $\delta$ and $u_0$ have been found to be related with temperature $T$ 
\cite{Kim:2007qk,Chakraborty:2017wdh} and $\Lambda_{\rm QCD}$ of gauge theory. Thus we give
a complete map of the non-AdS gravity and non-perturbative QCD. By expanding the dilaton near the boundary we obtain the form of gluon
condensate\footnote{The finite temperature gluon condensate has been studied previously in \cite{Kim:2007qk}. The gravity background
describing the gluon condensate at finite temperature has been identified, but the explicit form of gluon condensate and its temperature 
dependence (which we discuss in this Letter) has not been given there.} in this theory as a 
function of temperature.
The gluon condensate derived from this duality is found to have a form which resembles with that obtained
in SU(3) lattice QCD \cite{Miller:2006hr}. The condensate is found 
to disappear at the temperature where the background turns into a standard black D$3$ brane, i.e., in the deconfined phase. After that, 
using the renormalization group flow, we obtain the expression of the QCD beta function and plot it to show its energy dependence. The 
existence of non-trivial 
glueball mass is another characteristic of non-perturbative QCD. Using this gravity theory at zero temperature we holographically compute
both the pseudoscalar and scalar glueball mass spectra for the ground state and the first excited state at different gauge couplings.  
The mass ratios of the first excited state to the ground state are found to match with lattice result given in 
\cite{Morningstar:1999rf,FolcoCapossoli:2016ejd}. 

The decoupling limit of the non-supersymmetric (non-susy) D3 brane of type IIB string theory has been discussed in \cite{Nayek:2015tta,Nayek:2016hsi}.
The corresponding limit for the `black' non-susy D3 brane has been discussed in \cite{Chakraborty:2017wdh}. 
The decoupled geometry of the $N$ number of coincident `black' non-susy D$3$ brane solution is given in the string frame in eqs.(14) 
of ref.\cite{Chakraborty:2017wdh}. Here we consider that geometry with $\alpha+\beta=2$ in the Einstein frame (in this case, the radius 
of the S$^5$ part becomes constant and the computation becomes simpler) and is given by, 
\bea\label{geometry}
ds_{\rm Ein}^2 &=& \frac{u^2}{L^2}G(u)^{\frac{1}{4}-\frac{\d}{8}}\left(-G(u)^{\frac{\d}{2}} dt^2 
+ \sum_{i=1}^3 (dx^i)^2\right)\nn
& & +G(u)^{-1}\frac{L^2du^2}{u^2} + L^2 d\Omega_5^2\nn           
e^{2(\phi-\phi_0)} &=& G(u)^{\pm\half\sqrt{6 - \frac{3}{2}\delta^2}}\nn 
F_{[5]} &=& \frac{1}{\sqrt{2}}\left[1+\ast \right] 4L^4 {\rm Vol}(\Omega_5)\nn
{\rm with,}&& G(u) = 1 + \frac{u_0^4}{u^4}
\eea
Note that the solution is characterized by two free parameters $u_0$ and $\d$. It is clear from the expression of the dilaton in 
\eqref{geometry} that $|\d| \leq 2$. $L$ is the radius of the AdS$_5$ or S$^5$ space and is given by $L^4 = 2 N g_{\rm YM}^2 = 4 \pi g_s N
= \lambda_t$, where $g_{\rm YM}^2$ is the Yang-Mills coupling and $g_s=e^{\phi_0}$ is the string coupling which is assumed to be very small and
$\lambda_t$ is the 't Hooft coupling. Also in \eqref{geometry} `$\ast$' denotes the Hodge-dual operator and $F_{[5]}$ is self-dual.  
Now for the supergravity solution to remain valid the effective string coupling, $e^{\phi}$, must remain small and the radius of curvature 
of the transverse space in
string frame also must be large in unit of string length. In other words, we must have 
\bea
& & e^{\phi} \ll 1\nn
& & \frac{R^2}{\ell_s^2} = \sqrt{4\pi g_s N} e^{\frac{\phi-\phi_0}{2}} = \lambda_t e^{\frac{\phi-\phi_0}{2}} \gg 1
\eea
Here $\lambda_t\gg 1$ and $ds_\text{string}^2=e^{(\phi-\phi_0)/2}ds_\text{Ein}^2$. Now for the above two relations to hold, we observe
that $e^{\phi-\phi_0}$ must be of the order 1 and can never be $\ll 1$. Note that we are considering only the `$+$' sign in the exponent
of the dilaton expression\footnote{Here we are assuming that the parameter $|\delta| \neq 2$, because in that case $e^{\phi-\phi_0}$ is
independent of $u$. We will discuss $\delta = -2$ case later.} in \eqref{geometry} (as we will see later that this leads to the negativity 
of the beta function as in QCD)
and therefore, $e^{\phi - \phi_0}$ can not also be $\gg 1$. From the dilaton expression we therefore conclude that the energy parameter 
$u$ can never go to zero. Now, since the function $G(u)$ is always greater than or equal to
1, we have $e^{\phi - \phi_0} \geq 1$. Further, since $e^{\phi_0} \to 0$ and the 't Hooft coupling $\lambda_t \sim N e^{\phi_0} \gg 1$, it implies 
that $N \to \infty$ and so, the effective gauge theory coupling $\lambda \sim N e^{\phi}$ is also $\gg 1$, that is we are in the non-perturbative
regime of the gauge theory. Note that asymptotically as $u \to \infty$, $G(u) \to 1$ and the above solution reduces to the AdS$_5$ $\times$
S$^5$. This can also be obtained by taking $u_0 \to 0$. In this limit conformal symmetry will be restored and there is no QCD scale 
$\Lambda_{\rm QCD}$. It is, therefore, clear that non-zero $u_0$ must be proportional to that scale $\Lambda_{\rm QCD}$. To obtain a relation
between the two we notice that the gauge theory coupling is related to the dilaton as,         
\be\label{coupling}
\lambda = 4\pi N e^\phi=\lambda_t\left(1+\frac{u_0^4}{u^4}\right)^{\frac{1}{4}\sqrt{6-\frac{3}{2}\delta^2}}
\ee 
Now we find that at $u=\lambda_t^{\frac{1}{4}}u_0$, the harmonic function becomes $G(u) = 1 + \lambda_t^{-1} \approx 1$
and we get AdS$_5$, i.e., the conformal background and the coupling becomes $\lambda \approx \lambda_t$. Now we make an important assumption
that this energy where the theory becomes conformal defines the $\Lambda_{\rm QCD}$. Therefore, we have  
\be\label{lqcd}
u_0 = \lambda_t^{-\frac{1}{4}}\Lambda_{\rm QCD}
\ee 
Since $\lambda_t \gg 1$, it is clear that $u_0 \ll \Lambda_{\rm QCD}$. Now if the energy\footnote{Actually, the energy parameter $u$  
is related to the gauge theory energy $\kappa$ through the red-shift factor calculated in Einstein frame, as,
\begin{equation*}
\kappa=\frac{u}{\lambda_t^{1/4}}G(u)^{\frac{1}{8}+\frac{3}{16}\delta}.
\end{equation*}
One can show that $d\kappa/du > 0$ and so as $u$ increases $\kappa$ also increases. Therefore, for simplicity, we will take $u$ as the energy 
parameter of the gauge theory without any loss of generality.} 
$u$ is deep inside i.e., $u \ll \Lambda_\text{QCD}$, the ratio $u_0/u$ 
is of the order 1 and increases monotonically as $u$ further decreases. The coupling $\lambda$ also increases monotonically with the decrease of $u$. 
On the other hand, when 
$u$ is very close to $\Lambda_\text{QCD}$ the aforementioned ratio is of the order $\lambda_t^{-1/4}$ which is much smaller than 1. So in that limit 
one can write $\lambda=\lambda_t+\mathcal{O}(1/\lambda_t)$. Now if we take the energy beyond $\Lambda_\text{QCD}$ the effective gauge coupling 
becomes weaker but still it is slightly greater than $\lambda_t$. Thus 
for the whole range of the energy $0<u<\infty$, the effective gauge coupling $\lambda>\lambda_t\gg1$. So, the perturbative expansion in terms 
of $\lambda$ is not possible and the gauge theory remains in non-perturbative regime. Note that since we have here a energy dependent gauge coupling 
which decreases with the increase of the energy $u$ we can hope to have a non-perturbative QCD-like theory. 

The other parameter $\d$ can be shown to be related to the temperature of the background (or of the gauge theory) \cite{Chakraborty:2017wdh} by
\be\label{temperature}
T=\left(-\frac{\delta}{2}\right)^\frac{1}{4}\frac{u_0}{\pi\sqrt{\lambda_t}}
\ee
From this expression it is clear that the parameter $\d$ cannot be positive and so it must lie in the range $-2\leq \d \leq 0$. The temperature
can be made zero either by putting $\d=0$ or by putting $u_0=0$. In the first case, the solution reduces to the decoupled geometry of zero temperature
non-susy D3 brane discussed in \cite{Nayek:2016hsi}. In the second case, the parameter $\d$ disappears from the solution and it becomes the decoupled geometry of
standard BPS D3 brane, i.e., AdS$_5$ $\times$ S$^5$. Also for $\d=-2$, the solution reduces to the decoupled geometry of standard black D3 brane.

With this we proceed to obtain the expression for gluon condensate (GC) as a function of temperature. 
The detail method to calculate the GC from the holographic set-up has been given previously in \cite{Csaki:2006ji}. Here we are interested only in the temperature 
dependence of the GC. To find that dependence of the GC, we expand the dilaton field near the boundary $u\to\infty$.
\be\nonumber
\phi=\phi_0+\frac{1}{4}\sqrt{6-\frac{3}{2}\delta^2}\frac{u_0^4}{u^4}+{\mathcal O}(u^{-8})
\ee
As $e^{\phi_0},\,e^\phi\ll1$, $\phi_0$ and $\phi$ are negative quantities. The coefficient of $u^{-4}$ is identified as the GC, 
$\langle {\rm Tr}(G^2)\rangle$, at the finite temperature $T$\footnote{Assuming $\lambda_t=1.0$, this matches exactly with the zero temperature 
GC given in \cite{Csaki:2006ji}.}.
\be\label{gcondensate}
G_c(T)=\langle{\rm Tr} (G^2)\rangle=\frac{u_0^4}{4}\sqrt{6-\frac{3}{2}\delta^2}
\ee
Now replacing $\d$ in \eqref{gcondensate} from the temperature expression in \eqref{temperature} we get,
\be
G_c(T)=\frac{\sqrt{6}}{4}\pi^4\lambda_t^2\sqrt{\frac{u_0^8}{\pi^8\lambda_t^4}-T^8}
\ee
From the knowledge of QCD we know that, at $T=0$, GC have a finite nonzero value. Now as the temperature of the system is increased, 
the self coupling of the gluons decreases and as a result the value of the GC decreases. Finally at a certain temperature GC disappears - the temperature 
at that point is called the confinement temperature $T_c$. So $T_c$ can be easily found from the above expression by putting $G_c(T_c)=0$ and so, 
\be\label{tcritical}
T_c=\frac{u_0}{\pi\sqrt{\lambda_t}} ~~ =  \frac{\Lambda_\text{QCD}}{\pi\lambda_t^\frac{3}{4}}
\ee
The finite temperature gluon condensate $G_c(T)$, therefore, takes the following form
\bea\label{Gc}
G_c(T) & = & G_c(0)\sqrt{1-\frac{T^8}{T_c^8}}\label{gcfinal}\\
{\rm with}\quad G_c(0) & = & \frac{\sqrt{6}}{4}\pi^4\lambda_t^2T_c^4\nn
{\rm and}\quad\frac{T}{T_c} & = & \left(-\frac{\d}{2}\right)^\frac{1}{4}
\eea
At $T<T_c$, $G_c(T)\neq0$ indicates that the system is in confined state. Whereas at $T\geq T_c$, $G_c(T)=0$ indicates the system is in the deconfined state. 
So, $T_c$ is the transition temperature for the confinement-deconfinement transition\footnote{The transition temperature given in \eqref{tcritical} matches 
exactly with that derived by Herzog in \cite{Herzog:2006ra}.}. Eq.\eqref{tcritical} indicates that $T_c\ll \Lambda_\text{QCD}$. 
Also from \eqref{Gc} we note that the QCD scale $\Lambda_\text{QCD}$ can be related to the zero temperature GC by that relation.
\begin{figure}[h!]
\includegraphics[width=0.5\textwidth]{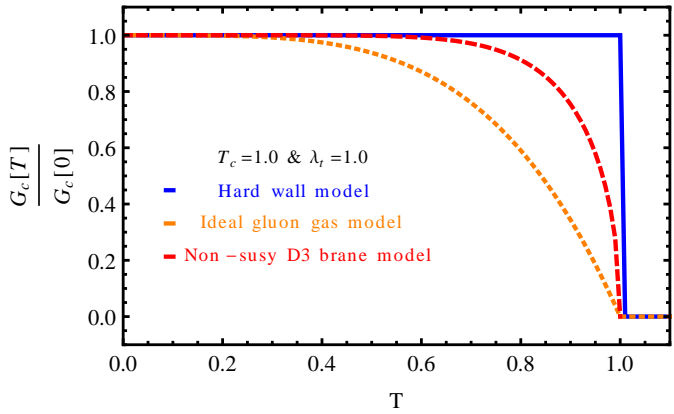}
\caption{Plots of normalized GC vs temperature for various models. The GC \eqref{gcfinal} calculated from non-susy D3 brane model is shown in red dashed line 
and has been compared with hard wall gravity model (blue solid line) and ideal gluon gas model (orange dotted line).}
\label{fig:gc}
\end{figure}
\begin{figure}[h!]
\includegraphics[width=0.5\textwidth]{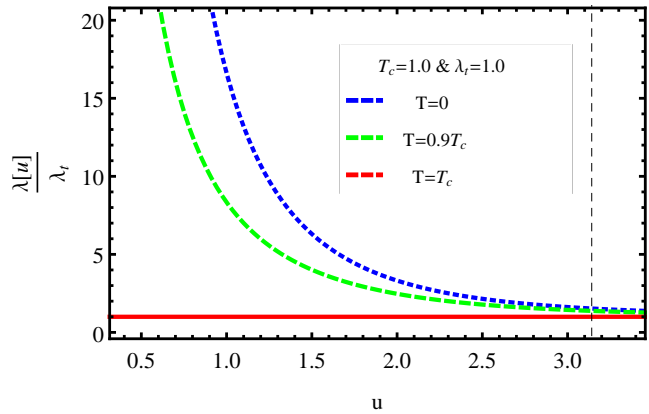}
\caption{Plots of normalized $\lambda$ vs $u$ given in \eqref{coupling1} for $T=0.0$ (blue dotted line), $0.9T_c$ (green dashed line), and $T_c$ (red solid line). 
Here $\lambda_t=1.0$ and $T_c=1.0$. The vertical dashed line indicates the position of $\Lambda_\text{QCD}=\pi$.}\label{fig:coupling1}
\end{figure}
In Fig.\ref{fig:gc}, we have plotted the gluon condensate given in \eqref{gcfinal}. In non-perturbative QCD, below the confinement scale there exists 
a non-trivial value of GC. It has a maximum value at $T=0$. As $T$ increases the interactive force of gluons decreases. Thus the GC decreases and vanishes 
at the confinement temperature $T_c$. All these have been shown in the figure. Here we have also compared our result with the hard wall dilaton-gravity model 
\cite{Kim:2007qk} and the ideal gluon gas condensate \cite{Miller:2006hr}. Our result falls in between those two results. But as we compare it with the 
lattice calculation, we see that our result is close to the GC derived in pure SU(3) lattice QCD \cite{Miller:2006hr}.

Now in order to compute $\b$-function we first express the effective coupling given in \eqref{coupling} in terms of the gauge theory parameters
as,
\be\label{coupling1}
\lambda  = \lambda_t\left(1+\frac{\Lambda_\text{QCD}^4}{\lambda_tu^4}\right)^{\frac{\sqrt{6}}{4}\sqrt{1-\frac{T^8}{T_c^8}}}
\ee
where we have used \eqref{tcritical} and \eqref{temperature}. Here the variations of coupling with energy $u$ and the temperature $T$ are clear. 
We note from \eqref{coupling1} that in the deep inside the IR regime $\lambda$ is very large, whereas, when $u \sim \Lambda_{\rm QCD}$, $\lambda \to \lambda_t$
(here $T < T_c$ is assumed).
The coupling also decreases with increasing temperature. At $T=T_c$ the coupling becomes constant $\lambda=\lambda_t$.  
In Fig.\ref{fig:coupling1} we have plotted $\lambda$ vs $u$. In numerical calculation, we have assumed $\lambda_t=1.0$ and $T_c=1.0$ which gives 
$\Lambda_\text{QCD}=\pi$. With these inputs we have numerically computed $\lambda$. It shows a monotonic decrease of coupling with increasing energy. At small 
energies the coupling decreases very fast and at high energies its variation is slower. As the energy is high enough, order of $\Lambda_{\rm QCD}$, coupling is 
very close to $\lambda_t$. Finally, at $u = \infty$ the coupling becomes constant, $\lambda_t$ and we move into the conformal gauge theory. The temperature 
variation of $\lambda$ is also shown in that same plot. It is found that at a fixed energy the gauge coupling decreases with increasing temperature. At the 
critical temperature $T_c$, the variation of $\lambda$ goes away and merges to $\lambda_t$.

The $\beta$-function can be calculated from the coupling $\lambda$, \eqref{coupling1}, by using the renormalization group flow equation as follows,
\bea
\beta & = & \frac{d\lambda}{d\ln u}\nn
& = & -\sqrt{6}\frac{\Lambda_\text{QCD}^4}{u^4}\sqrt{1-\frac{T^8}{T_c^8}}\left(1+\frac{\Lambda_\text{QCD}^4}{\lambda_tu^4}\right)^{\frac{\sqrt{6}}{4}\sqrt{1-\frac{T^8}{T_c^8}}-1}\nn
& = & -\sqrt{6}\lambda\sqrt{1-\frac{T^8}{T_c^8}}\left[1-\left(\frac{\lambda_t}{\lambda}\right)^\frac{4T_c^4}{\sqrt{6}\sqrt{T_c^8-T^8}}\right]
\eea 
We have expressed the $\beta$-function as functions of $u$ and $\lambda$ separately. The non-zero value of $\beta$ indicates that the theory has a running coupling 
and its negativity indicates that $\lambda$ decreases with energy $u$. We have $\beta(u \to 0)\to-\infty$, $\beta(u\to\infty)\to0$ and at $u=\Lambda_\text{QCD}$, 
$\beta=-\sqrt{6}\sqrt{1-\frac{T^8}{T_c^8}}$ takes a negative constant value. 
In terms of $\lambda$, at high energies when $\lambda\to\lambda_t$, $\beta$ goes to zero. But as we move towards the deep IR regime, 
$\lambda\gg\lambda_t$, $\beta$ takes high negative value and finally as $\lambda\to\infty$, $\beta\to-\infty$. 
\begin{figure}[h!]
\includegraphics[width=0.5\textwidth]{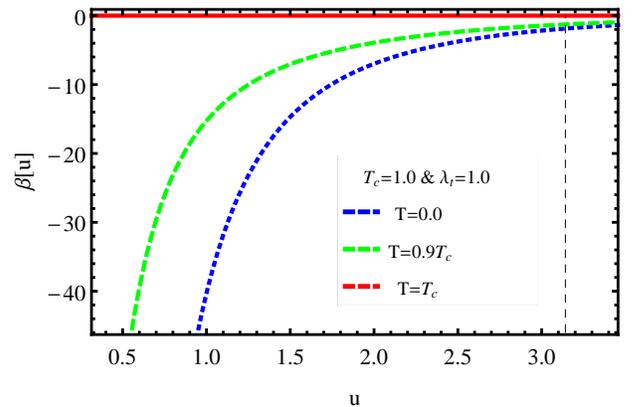}
\caption{Plots of $\beta$ vs $u$ for $T=0.0$ (blue dotted line), $0.9T_c$ (green dashed line), and $T_c$ (red solid line). Here $\lambda_t=1.0$ and $T_c=1.0$. 
The vertical dashed line indicates the position of $\Lambda_\text{QCD}=\pi$.}\label{fig:beta}
\end{figure}
In Fig.\ref{fig:beta}, we have shown the variation of $\beta$-function with energy and temperature. Here $\beta$ is always negative that means the coupling 
is a monotonically decreasing function of energy. At small $u$, the higher values of $\beta$ indicate the faster variation of $\lambda$. Again at low temperature, 
$\beta$ takes more negative value indicating the rapid variation of coupling $\lambda$. For $T=T_c$, $\beta$ is zero which is consistent with the fact that at 
deconfined temperature coupling becomes constant. For other finite $T$, $\beta$ merges to $-\sqrt{6}\sqrt{1-T^8/T_c^8}$ near $u=\Lambda_\text{QCD}$. 
In $u\ge\Lambda_\text{QCD}$ regime, $\lambda$ is almost same for all $T$ but $\beta$ function differ with a finite value depending on $T$. This difference 
is easily visible in Fig.\ref{fig:beta} but not in Fig.\ref{fig:coupling1}.  

The existence of non-trivial glueball mass is another low energy or non-perturbative property of QCD. Although experimentally, the glueballs
have not been observed, theoretically the glueball mass spectra has been calculated, particularly using 
lattice QCD \cite{Morningstar:1999rf,FolcoCapossoli:2016ejd}. The same results have been obtained from the holographic QCD approaches 
\cite{Csaki:1998qr,deMelloKoch:1998vqw,Minahan:1998tm,Constable:1999ch}. Here in this work we use 
the gravity/QCD correspondence for the decoupled geometry of non-susy D$3$ brane solution of type-IIB string theory to compute the glueball mass spectra.
We compare our results with some lattice results \cite{Morningstar:1999rf,FolcoCapossoli:2016ejd} and found good agreement. 
Here we will focus only on the masses of the ground state and the first excited state of spin-$0$ scalar and pseudoscalar glueballs $0^{++},\,0^{-+}$. 
To obtain the glueball masses we consider the linearized equations of motion for the scalar field fluctuations corresponding 
to the dilaton and the axion propagating in the decoupled non-susy D3 brane background. The Schr\" odinger-like equation 
of these scalar fluctuations obtained in this way has been solved using the WKB approximation \cite{Constable:1999ch,Minahan:1998tm} to obtain the mass spectra. 
Unlike in the previous case, here we consider only the zero temperature solution which can be obtained from eqn (14) of ref.\cite{Chakraborty:2017wdh} 
with $\delta=0$, but we keep the other parameters arbitrary. Here we put $\alpha+\beta=\gamma$ (instead of 2 as in the previous case), which is
not fixed. The background solution in the string frame is given below\footnote{This zero temperature, decoupled non-susy D3 brane solution of type
IIB string theory has been shown before by us \cite{Nayek:2016hsi} to be identical with the solution obtained by Constable and Myers in 
\cite{Constable:1999ch} by a coordinate transformation
and redefinition of the parameters. In their paper, they also discussed the issue of mass gap and glueball mass spectra in the boundary gauge theory. They 
have obtained the form of mass spectra analytically by solving the WKB equation in the large mass limit. We solve the equation numerically without
assuming the mass to be large and obtain the numerical values of the masses in certain units. We obtain both the scalar and pseudoscalar glueball masses,
whereas in \cite{Constable:1999ch} only the pseudoscalar glueball mass formula has been given and mentioned that the scalar glueball mass can not be obtained using
supergravity. We, however, do not face such problem in our computation. More discussion will be given towards the end.},
\bea\label{background}
ds^2 & = & \sqrt{\frac{\gamma u_0^4}{2L^4}}\frac{G(u)^\frac{\delta_1}{4}}{\sqrt{F(u)}}\left(-dt^2+\sum_{i=1}^3(dx^i)^2\right)\nn
&& + \sqrt{\frac{2L^4}{\gamma u_0^4}}\sqrt{F(u)}G(u)^{\frac{2+5\delta_1}{8}}\left(\frac{du^2}{G(u)}+u^2d\Omega_5^2\right)\nn
e^{2\phi} &=& g_s^2G(u)^\frac{7\delta_1}{4}\nn
F_5 &=& \frac{1}{\sqrt{2}}[1+\ast]4L^4\text{vol}(\Omega_5)
\eea
where, 
\bea
F(u) & = & G(u)^\frac{\alpha}{2}-G(u)^{-\frac{\beta}{2}}\nn
\alpha - \beta & = & -\frac{3}{2}\delta_1\nn
\alpha+\beta=\gamma &=& \sqrt{10-\frac{49}{4}\delta_1^2}\nonumber
\eea
The background has explicitly two parameters which are $u_0$ and $\delta_1$. In dual gauge theory $u_0$ is related to the QCD scale 
$\Lambda_{\rm QCD}$ and $\d_1$ determines the form of the coupling and so, different $\d_1$'s give different gauge theories as the form of
the coupling $e^{\phi}$ changes with $\d_1$.
We assume $\varphi$ and $\chi$ to be the fluctuations of the dilaton and the axion respectively. So the linearized equation of the dilaton fluctuation 
$\varphi$ in the Einstein frame is 
\be\label{dilaton_fluc}
\partial_\mu\left(\sqrt{-g}g^{\mu\nu}\partial_\nu\right)\varphi=0
\ee
and the linearized equation of the axion fluctuation $\chi$ in the string frame is
\be\label{axion_fluc}
\partial_\mu\left(\sqrt{-g}g^{\mu\nu}\partial_\nu\right)\chi=0
\ee
In case of the dilaton fluctuation, the background metric is also perturbed. But using a particular gauge condition the metric fluctuation can be eliminated
from the linearized dilaton equation and we have written \eqref{dilaton_fluc} in that particular gauge. So, even though the metric fluctuation is present, 
there is no need to solve the 
full linearized Einstein's equation if we choose the suitable gauge \cite{Constable:1999ch}.
Now as we are in the near-horizon or decoupled geometry, we can take $\varphi$ to be symmetric in the transverse $\Omega_5$ direction. In other words, 
we take $\varphi=h(u)e^{i{\bf p.x}}$ and mass of this glueball $0^{++}$ is $M$ where $p_\mu p^\mu=-M^2$. Using this particular form of the fluctuation and the 
background \eqref{background} into the \eqref{dilaton_fluc}, we get the Schr\" odinger-like wave equation in Einstein frame as,
\be\label{wave_eqn}
\psi''(y)-V_{\rm dilaton}(y)\psi(y)=0
\ee
where $\psi(y)=e^{2y}G(y)^{\frac{1}{2}+\frac{7}{16}\delta_1}h(y)$ and the potential is 
\bea\label{potential}
&V_\text{dilaton}(y) = \frac{49}{16}\frac{\delta_1^2}{(1+e^{4y})^2}+4\frac{1+2e^{-4y}}{(1+e^{-4y})^2}&\nonumber\\
& -\frac{2L^4M^2}{\gamma u_0^2}e^{2y}\left[(1+e^{-4y})^\frac{\gamma-3}{4}-(1+e^{-4y})^{-\frac{\gamma+3}{4}}\right]&
\eea
Note that the above equation \eqref{wave_eqn} is written in a new variable $y$ which is related to $u$ by the relation $y = \ln(u/u_0)$.
The potential \eqref{potential} gives a little potential well with two boundaries or turning points on the two sides around $y=0$.  
For $y\gg0$,
\be
V_{\rm dilaton}\approx 4-\frac{L^4M^2}{u_0^2}e^{-2y}.
\ee 
So, the positive turning point is $y_+=\ln\left(\frac{L^2M}{2u_0}\right)$ and for $y\ll0$,
\be\label{dilatonpot}
V_{\rm dilaton}\approx \frac{49}{16}\delta_1^2-\frac{2L^4M^2}{\gamma u_0^2}e^{(5-\gamma)y}
\ee
and the negative turning point is at $y_-=\frac{1}{5-\gamma}\ln\left(\frac{49\gamma}{32M^2}\delta_1^2\right)$. Now according to the WKB 
approximation, if the 
depth of the potential well is very small, the mass of the $n$-th excited state can be found from the following equality.
\be\label{wkb}
\left(n-\frac{1}{2}\right)\pi=\int_{y_-}^{y_+}\sqrt{-V(y)}dy, \qquad {\rm for,}\quad n=1,2,3,\ldots
\ee
Now in the above equality, the right hand side is a function of $M$ and so the mass $M$ can be easily found by solving the algebraic equation of $M$. Here $n=1$ 
gives ground-state mass, $n=2$ gives mass of the first excited state and so on. We follow the same procedure for the axion fluctuation.
In this case , the potential is the same as \eqref{potential} except the first term. The axion fluctuation potential is then given as follows,
\bea\label{axionpot}
&V_\text{axion}(y) = \frac{49}{4}\frac{\delta_1^2}{(1+e^{4y})^2}+4\frac{1+2e^{-4y}}{(1+e^{-4y})^2}&\nonumber\\
&-\frac{2L^4M^2}{\gamma u_0^2}e^{2y}\left[(1+e^{-4y})^\frac{\gamma-3}{4}-(1+e^{-4y})^{-\frac{\gamma+3}{4}}\right]&
\eea
Here the positive turning point is the same as in the dilaton case, but the negative turning point is 
$y_-= \frac{1}{5-\gamma}\ln\left(\frac{49\gamma}{8M^2}\delta_1^2\right)$.   
In numerical calculation we always deal with dimensionless quantities. So here the masses are found in the units of $u_0/L^2$. Various lattice 
computations \cite{Morningstar:1999rf} also have given this spectrum in their own units. To eliminate this ambiguity we compare the ratio of masses 
of the above mentioned two states and are given in the table below.    
\begin{table}[h!]
\caption{Here the glueball masses are given in units of $\frac{u_0}{L^2}$ MeV or in units of $\frac{\Lambda_\text{QCD}}{\lambda_t^{\frac{3}{4}}}$ MeV.}
\begin{tabular}{|c|c|c|c|c|c|c|c|}
\hline\hline
$\delta_1$ & $M_{0^{++}}$ & $M_{0^{++*}}$ & $\frac{M_{0^{++*}}}{M_{0^{++}}}$& $M_{0^{-+}}$ & $M_{0^{-+*}}$ & $\frac{M_{0^{-+*}}}{M_{0^{-+}}}$ & $\frac{M_{0^{-+}}}{M_{0^{++}}}$ \\
\hline
0.0	&	2.5788	&	4.4213	&	1.7145	&	2.5788	&	4.4214	&	1.7145	&	1.0000\\
\hline
0.05	&	2.6825	&	4.5123	&	1.6822	&	2.7768	&	4.5948	&	1.6547	&	1.0352\\
\hline
0.1	&	2.7852	&	4.6077	&	1.6544	&	2.9566	&	4.7692	&	1.6131	&	1.0615\\
\hline
0.2	&	2.9855	&	4.8171	&	1.6135	&	3.2810	&	5.1073	&	1.5566	&	1.0990\\
\hline
0.3	&	3.1800	&	5.0313	&	1.5822	&	3.5729	&	5.4630	&	1.5290	&	1.1236\\
\hline
0.4	&	3.3591	&	5.2805	&	1.5720	&	3.8203	&	5.7900	&	1.5156	&	1.1373\\
\hline
0.5	&	3.5475	&	5.5048	&	1.5517	&	4.0300	&	6.0924	&	1.5118	&	1.1360\\
\hline
0.6	&	3.7088	&	5.7593	&	1.5529	&	4.2081	&	6.3674	&	1.5131	&	1.1346\\
\hline
0.7	&	3.8566	&	5.9850	&	1.5519	&	4.3617	&	6.6194	&	1.5176	&	1.1310\\
\hline
0.8	&	3.9888	&	6.1936	&	1.5528	&	4.4972	&	6.8400	&	1.5209	&	1.1275\\
\hline
0.9	&	4.1032	&	6.3898	&	1.5573	&	4.6188	&	7.0465	&	1.5256	&	1.1257\\
\hline\hline
\end{tabular}
\end{table}

Here we have shown the masses of the ground state and the first excited state of the scalar and the pseudoscalar glueballs at various values of
the parameter $\d_1$ and also the ratios of the masses of the first excited state and the ground state for both the scalar and the pseudoscalar
glueballs. In the final column we have shown the ratios of the masses of the ground state scalar and pseudoscalar glueballs. We see that
when $\d_1=0$, both the scalar and the pseudoscalar glueballs have the same masses. The reason is for $\d_1=0$, the dilaton becomes constant and
therefore, the fluctuation equations for both the dilaton and the axion become identical and give identical solutions. Different $\d_1$ actually
defines different theories with different gauge couplings. So, the glueball masses are also different for different $\d_1$. We notice that the
the ratios of masses of the first excited state and the ground state for the scalar glueball varies from 1.7145 to 1.5517. The average value of the 
scalar glueball masses obtained from lattice calculation by various groups are listed in \cite{FolcoCapossoli:2016ejd} and from there we find 
that the ratio of the
mass of the first excited state to the ground state of the scalar glueball takes the value $0^{++\ast}/0^{++} = 2.751/1.595 = 1.725$. So, it is 
quite close to the results we obtain from the decoupled non-susy D3-brane geometry. Also from the last column of the above table we notice
that the mass of the pseudoscalar glueball is greater than the mass of the scalar glueball but the difference is not much since the ratio is close to 1.
If we look at the lattice results, again from the average values obtained by various groups we find the ratio $0^{+-}/0^{++} = 1.595/2.467 = 1.547$.
Here our result differs and this could be due to the fact that our results are valid only at strong coupling. In fact, in \cite{Tsue:2012kf}, using
some time-dependent variational approach it has been claimed that at strong coupling the mass ratios of the pseudoscalar and the scalar glueballs
in Yang-Mills theory must tend to 1.

We remark that in obtaining the glueball masses, we have to perform an integration \eqref{wkb} with the integration limits from $y_-$ to
$y_+$, the two turning points of the potential. For both scalar and pseudoscalar glueball masses, the positive turning points $y_+$ are the same
and fixed (depends only on mass $M$). On the other hand the negative turning points $y_-$ for both the cases depend on $\delta_1$, $\gamma$ and 
$M$ (see the expressions for $y_-$ after \eqref{dilatonpot} and \eqref{axionpot}). There are three cases where the computation could be
problematic (i) $\delta_1 \to 0$, (ii) $\gamma \to 0$ and (iii) $M^2 \to \infty$. In all three cases $y_- \to -\infty$ and we have to check whether
the supergravity description remains valid there. For all other cases there are no problem. We notice that for case (i) when $\d_1 \to 0$ the
dilaton goes to constant and therefore $y_- \to -\infty$ does not pose any problem and supergravity description remains valid there. Notice from 
\eqref{background} that $\d_1 \geq 0$ and lies in the range $0\leq \d_1 \leq \sqrt{40/49} \approx 0.904$. Also we mention that $\d_1 \to 0$
corresponds to $\Delta \to 0$ in \cite{Constable:1999ch} and in this case the string frame and the Einstein frame metrics coincide and 
therefore we get the same
equation for the dilaton and axion fluctuations. The masses are given in the first line of Table 1. $\gamma$ on the other hand lies in the
range $0 \leq \gamma \leq \sqrt{10}$, where $\gamma=0$ corresponds to $\d_1 = \sqrt{40/49}$ and $\gamma = \sqrt{10}$ corresponds to $\d_1=0$.
For case (ii), when $\gamma \to 0$, $\d_1$ remains finite there and so $y_- \to -\infty$ makes $e^{\phi-\phi_0}$ to blow up and therefore
supergravity description breaks down. In other words we can not trust the mass calculation near $\gamma=0$. For case (iii), when $M^2 \to \infty$
which means that we are considering highly excited states, again if $\d_1$ does not vanish, the dilaton blows up making the supergravity
description invalid and we cannot trust the glueball mass calculation. In all other cases, glueball masses we obtain are quite reliable
indicating the theory possesses a mass gap like QCD.            

In this Letter, we have studied some non-perturbative aspects of QCD by making use of the gravity/QCD type correspondence applied to
the decoupled geometry of non-susy D3 brane of type IIB string theory. Since the gravity theory here is non-supersymmetric, non-conformal
and has non-trivial dilaton, the corresponding boundary theory is more like QCD. The background also has a temperature and so
the QCD-like theory is at finite temperature. We have obtained the form of gluon condensate and expressed it as a function of temperature.
We plotted the gluon condensate versus temperature and found that the form is close to that found in lattice calculation. We also obtained
the expression of the gauge coupling and also the beta function from the renormalization group flow equation. Beta function is found to
be negative as in QCD. We have plotted both the gauge coupling and the beta function and discussed their behavior. Finally, to study other
non-perturbative aspects of QCD, we have computed both the scalar and the pseudoscalar glueball mass spectra in our theory obtained
from non-susy D3 brane and compared with the lattice results.



\begin{thebibliography}{99}

\bibitem{Maldacena:1997re} 
  J.~M.~Maldacena,
  ``The Large N limit of superconformal field theories and supergravity,''
  Int.\ J.\ Theor.\ Phys.\  {\bf 38}, 1113 (1999)
  [Adv.\ Theor.\ Math.\ Phys.\  {\bf 2}, 231 (1998)]
  doi:10.1023/A:1026654312961, 10.4310/ATMP.1998.v2.n2.a1
  [hep-th/9711200].

\bibitem{Aharony:1999ti} 
  O.~Aharony, S.~S.~Gubser, J.~M.~Maldacena, H.~Ooguri and Y.~Oz,
  ``Large N field theories, string theory and gravity,''
  Phys.\ Rept.\  {\bf 323}, 183 (2000)
  doi:10.1016/S0370-1573(99)00083-6
  [hep-th/9905111].

\bibitem{Polchinski:2001tt} 
  J.~Polchinski and M.~J.~Strassler,
  ``Hard scattering and gauge / string duality,''
  Phys.\ Rev.\ Lett.\  {\bf 88}, 031601 (2002)
  doi:10.1103/PhysRevLett.88.031601
  [hep-th/0109174];
  J.~Polchinski and M.~J.~Strassler,
  ``Deep inelastic scattering and gauge / string duality,''
  JHEP {\bf 0305}, 012 (2003)
  doi:10.1088/1126-6708/2003/05/012
  [hep-th/0209211].

\bibitem{Constable:1999ch} 
  N.~R.~Constable and R.~C.~Myers,
  ``Exotic scalar states in the AdS / CFT correspondence,''
  JHEP {\bf 9911}, 020 (1999)
  doi:10.1088/1126-6708/1999/11/020
  [hep-th/9905081].

\bibitem{Babington:2003vm} 
  J.~Babington, J.~Erdmenger, N.~J.~Evans, Z.~Guralnik and I.~Kirsch,
  ``Chiral symmetry breaking and pions in nonsupersymmetric gauge / gravity duals,''
  Phys.\ Rev.\ D {\bf 69}, 066007 (2004)
  doi:10.1103/PhysRevD.69.066007
  [hep-th/0306018].

\bibitem{Csaki:2006ji} 
  C.~Csaki and M.~Reece,
  ``Toward a systematic holographic QCD: A Braneless approach,''
  JHEP {\bf 0705}, 062 (2007)
  doi:10.1088/1126-6708/2007/05/062
  [hep-ph/0608266].

\bibitem{Liu:2006he} 
  H.~Liu, K.~Rajagopal and U.~A.~Wiedemann,
  ``Wilson loops in heavy ion collisions and their calculation in AdS/CFT,''
  JHEP {\bf 0703}, 066 (2007)
  doi:10.1088/1126-6708/2007/03/066
  [hep-ph/0612168].

\bibitem{CasalderreySolana:2011us} 
  J.~Casalderrey-Solana, H.~Liu, D.~Mateos, K.~Rajagopal and U.~A.~Wiedemann,
  ``Gauge/String Duality, Hot QCD and Heavy Ion Collisions,''
  book:Gauge/String Duality, Hot QCD and Heavy Ion Collisions. Cambridge, UK: Cambridge University Press, 2014
  doi:10.1017/CBO9781139136747
  [arXiv:1101.0618 [hep-th]].

\bibitem{Chakraborty:2017wdh} 
  S.~Chakraborty, K.~Nayek and S.~Roy,
  ``Wilson loop calculation in QGP using non-supersymmetric AdS/CFT,''
  arXiv:1710.08631 [hep-th].

\bibitem{Batell:2008zm} 
  B.~Batell and T.~Gherghetta,
  ``Dynamical Soft-Wall AdS/QCD,''
  Phys.\ Rev.\ D {\bf 78}, 026002 (2008)
  doi:10.1103/PhysRevD.78.026002
  [arXiv:0801.4383 [hep-ph]].

\bibitem{Gubser:1999pk} 
  S.~S.~Gubser,
  ``Dilaton driven confinement,''
  hep-th/9902155.

\bibitem{Kehagias:1999tr} 
  A.~Kehagias and K.~Sfetsos,
  ``On Running couplings in gauge theories from type IIB supergravity,''
  Phys.\ Lett.\ B {\bf 454}, 270 (1999)
  doi:10.1016/S0370-2693(99)00393-7
  [hep-th/9902125].

\bibitem{Nojiri:1999uh} 
  S.~Nojiri and S.~D.~Odintsov,
  ``Running gauge coupling and quark - anti-quark potential in nonSUSY gauge theory at finite temperature from IIB SG / CFT correspondence,''
  Phys.\ Rev.\ D {\bf 61}, 024027 (2000)
  doi:10.1103/PhysRevD.61.024027
  [hep-th/9906216].

\bibitem{Brodsky:2010ur} 
  S.~J.~Brodsky, G.~F.~de Teramond and A.~Deur,
  ``Nonperturbative QCD Coupling and its $\beta$-function from Light-Front Holography,''
  Phys.\ Rev.\ D {\bf 81}, 096010 (2010)
  doi:10.1103/PhysRevD.81.096010
  [arXiv:1002.3948 [hep-ph]].

\bibitem{deTeramond:2009xk} 
  G.~F.~de Teramond and S.~J.~Brodsky,
  ``Light-Front Holography and Gauge/Gravity Duality: The Light Meson and Baryon Spectra,''
  Nucl.\ Phys.\ Proc.\ Suppl.\  {\bf 199}, 89 (2010)
  doi:10.1016/j.nuclphysbps.2010.02.010
  [arXiv:0909.3900 [hep-ph]].

\bibitem{Zhou:1999nm} 
  B.~Zhou and C.~J.~Zhu,
  ``The Complete black brane solutions in D-dimensional coupled gravity system,''
  hep-th/9905146.

\bibitem{Lu:2004ms} 
  J.~X.~Lu and S.~Roy,
  ``Static, non-SUSY p-branes in diverse dimensions,''
  JHEP {\bf 0502}, 001 (2005)
  doi:10.1088/1126-6708/2005/02/001
  [hep-th/0408242].

\bibitem{Nayek:2015tta} 
  K.~Nayek and S.~Roy,
  ``Decoupling of gravity on non-susy D$p$ branes,''
  JHEP {\bf 1603}, 102 (2016)
  doi:10.1007/JHEP03(2016)102
  [arXiv:1506.08583 [hep-th]].

\bibitem{Nayek:2016hsi} 
  K.~Nayek and S.~Roy,
  ``Decoupling limit and throat geometry of non-susy D3 brane,''
  Phys.\ Lett.\ B {\bf 766}, 192 (2017)
  doi:10.1016/j.physletb.2017.01.007
  [arXiv:1608.05036 [hep-th]].

\bibitem{Kim:2007qk} 
  Y.~Kim, B.~H.~Lee, C.~Park and S.~J.~Sin,
  ``Gluon Condensation at Finite Temperature via AdS/CFT,''
  JHEP {\bf 0709}, 105 (2007)
  doi:10.1088/1126-6708/2007/09/105
  [hep-th/0702131].

\bibitem{Miller:2006hr} 
  D.~E.~Miller,
  ``Lattice QCD Calculation for the Physical Equation of State,''
  Phys.\ Rept.\  {\bf 443}, 55 (2007)
  doi:10.1016/j.physrep.2007.02.012
  [hep-ph/0608234].

\bibitem{Morningstar:1999rf} 
  C.~J.~Morningstar and M.~J.~Peardon,
  ``The Glueball spectrum from an anisotropic lattice study,''
  Phys.\ Rev.\ D {\bf 60}, 034509 (1999)
  doi:10.1103/PhysRevD.60.034509
  [hep-lat/9901004];
  B.~Lucini and M.~Teper,
  ``SU(N) gauge theories in four-dimensions: Exploring the approach to N = infinity,''
  JHEP {\bf 0106}, 050 (2001)
  doi:10.1088/1126-6708/2001/06/050
  [hep-lat/0103027];
  H.~B.~Meyer,
  ``Glueball regge trajectories,''
  hep-lat/0508002;
  Y.~Chen {\it et al.},
  ``Glueball spectrum and matrix elements on anisotropic lattices,''
  Phys.\ Rev.\ D {\bf 73}, 014516 (2006)
  doi:10.1103/PhysRevD.73.014516
  [hep-lat/0510074].

\bibitem{FolcoCapossoli:2016ejd} 
  D.~M.~Rodrigues, E.~Folco Capossoli and H.~Boschi-Filho,
  ``Scalar and higher even spin glueball masses from an anomalous modified holographic model,''
  EPL {\bf 122}, no. 2, 21001 (2018)
  doi:10.1209/0295-5075/122/21001
  [arXiv:1611.09817 [hep-ph]].

\bibitem{Herzog:2006ra} 
  C.~P.~Herzog,
  ``A Holographic Prediction of the Deconfinement Temperature,''
  Phys.\ Rev.\ Lett.\  {\bf 98}, 091601 (2007)
  doi:10.1103/PhysRevLett.98.091601
  [hep-th/0608151].

\bibitem{Csaki:1998qr} 
  C.~Csaki, H.~Ooguri, Y.~Oz and J.~Terning,
  ``Glueball mass spectrum from supergravity,''
  JHEP {\bf 9901}, 017 (1999)
  doi:10.1088/1126-6708/1999/01/017
  [hep-th/9806021].

\bibitem{deMelloKoch:1998vqw} 
  R.~de Mello Koch, A.~Jevicki, M.~Mihailescu and J.~P.~Nunes,
  ``Evaluation of glueball masses from supergravity,''
  Phys.\ Rev.\ D {\bf 58}, 105009 (1998)
  doi:10.1103/PhysRevD.58.105009
  [hep-th/9806125].

\bibitem{Minahan:1998tm} 
  J.~A.~Minahan,
  ``Glueball mass spectra and other issues for supergravity duals of QCD models,''
  JHEP {\bf 9901}, 020 (1999)
  doi:10.1088/1126-6708/1999/01/020
  [hep-th/9811156].

\bibitem{Tsue:2012kf} 
  Y.~Tsue,
  ``Scalar and Pseudoscalar Glueball Masses within a Gaussian Wavefunctional Approximation,''
  Prog.\ Theor.\ Phys.\  {\bf 128}, 373 (2012)
  doi:10.1143/PTP.128.373
  [arXiv:1204.0296 [hep-ph]].

\end{thebibliography}
\end{document}